\documentclass[a4paper, 10pt, conference]{ieeeconf}      

\IEEEoverridecommandlockouts                              

\usepackage{graphicx}
\usepackage{amsmath}
\usepackage[czech]{babel}

\usepackage[markup=bfit, deletedmarkup=sout, authormarkup=brackets]{changes}
\definechangesauthor[name={Karla}, color=teal]{KS}
\definechangesauthor[name={Michal}, color=orange]{MV}
\definechangesauthor[name={Jozin}, color=red]{JS}
\definechangesauthor[name={Robert}, color=brown]{RB}
\definechangesauthor[name={Rado}, color=purple]{RS}
\definechangesauthor[name={Jirka}, color=pink]{JI}
\definechangesauthor[name={Gabina}, color=cyan]{GS}
\definechangesauthor[name={Michael}, color=green]{MT}

\title{\LARGE \bf
Teaching robots to imitate a human with no on-teacher sensors. What are the key challenges?
}

\author{Radoslav Skoviera$^{1}$, Karla Stepanova$^{1}$, Michael Tesar$^{1}$, Gabriela Sejnova$^{1}$, Jiri Sedlar$^{1}$, Michal Vavrecka$^{1}$, \\
Robert Babuska$^{1,2}$, and Josef Sivic$^{1,3}$
\thanks{$^{1}$Czech Institute of Informatics, Robotics, and Cybernetics, Czech Technical University, Prague, Czech Republic}%
\thanks{$^{2}$TU Delft}%
\thanks{$^{3}$Inria, WILLOW, Departement d'Informatique de l'Ecole Normale Superieure, PSL Research University, ENS/INRIA/CNRS UMR 8548, Paris, France}%
}

\begin{document}

\maketitle
\thispagestyle{empty}
\pagestyle{empty}

\begin{abstract}
In this paper, we consider the problem of learning object manipulation tasks from human demonstration using RGB or RGB-D cameras. We highlight the key challenges in capturing sufficiently good data with no tracking devices -- starting from sensor selection and accurate 6DoF pose estimation to natural language processing. In particular, we focus on two showcases: gluing task with a glue gun and simple block-stacking with variable blocks. Furthermore, we discuss how a linguistic description of the task could help to improve the accuracy of task description. We also present the whole architecture of our transfer of the imitated task to the simulated and real robot environment.

\end{abstract}


\section{INTRODUCTION}

Imitation learning has a long history and many applications ranging from easier programming of industrial robots to household companions. Nonetheless, in practice, it still heavily relies on tracking devices and optical motion tracking~\cite{guerra2005optitrack} accompanied by markers on objects which help to identify 6D pose of the given object.
If we want to develop social and cognitive robots which are able to learn from a direct interaction with humans and imitate or recognize their actions, we have to develop teaching methods which will not require any on-teacher sensors and will be able to learn in a natural environment (e.g., household). These robots will have to learn from direct observation (visual, linguistic, haptic) using only their own sensors and previously acquired knowledge. Our biggest concern is, whether current hardware and machine learning methods enable this type of imitation learning. Can recent rapid progress in computer vision and natural language processing techniques make omitting cumbersome, task-specific sensing devices possible and enable robots to really understand the scene and reality they are observing?  

\begin{figure}[thpb]
\centering
\framebox{\parbox{3.2in}{\includegraphics[width=230 pt]{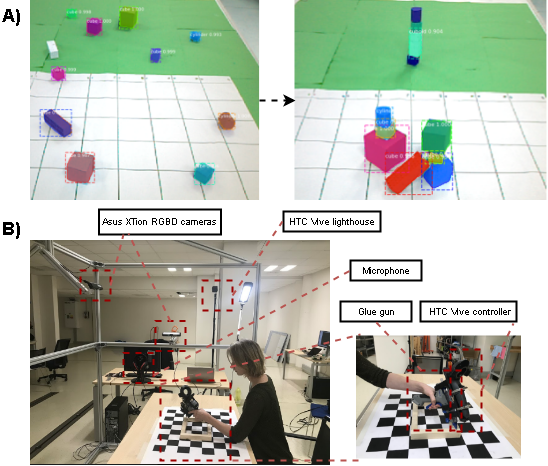}}}
\caption{A) Showcase 1: a simple block-stacking tasks with variable blocks, B) Training center with various sensors; Showcase 2: glue application with a hot glue gun.}
\label{fig:setup}
\end{figure}

Imitation learning first became an object of interest in the early 1980s as a possible way towards higher autonomy in industrial robots. The initial approach was manual operation of the robot, such as the \textit{teach-in} method,\textit{ guiding} or a \textit{play-back} method. The demonstrated task was represented as a series of transitions between states and actions, which were further converted into a set of graph-based symbolic rules and relationships~\cite{billard2008robot}.    
Ever since, most of the progress happened in the field of teaching methods, varying from vision to kinesthetic teaching, where the robot is physically manipulated to perform the desired task. Remote teaching typically includes various on-teacher sensors. Argall~et~al.~\cite{argall2010} tried to use only sensors on the robot to mimic human behavior. There are also some attempts to detect human activity from RGB-D sensors \cite{zimmermann20183d} as well as from RGB narrated videos \cite{alayrac2016}. Visual demonstration accompanied by language instructions was also used in~\cite{liu2016} where they tried to learn grounded task structures for T-shirts folding task. M{\"u}hlig et al.~\cite{muhlig2012} tried to teach a robot manipulation block-stacking task from a tutor sitting behind a table. 
To our best knowledge, there is no work done on learning from demonstration relying only on visual, linguistic and haptic information without any on-teacher sensors for complex manipulation tasks in a real-world environment.

\begin{figure*}[thpb]
\centering
{\includegraphics[width= \textwidth]{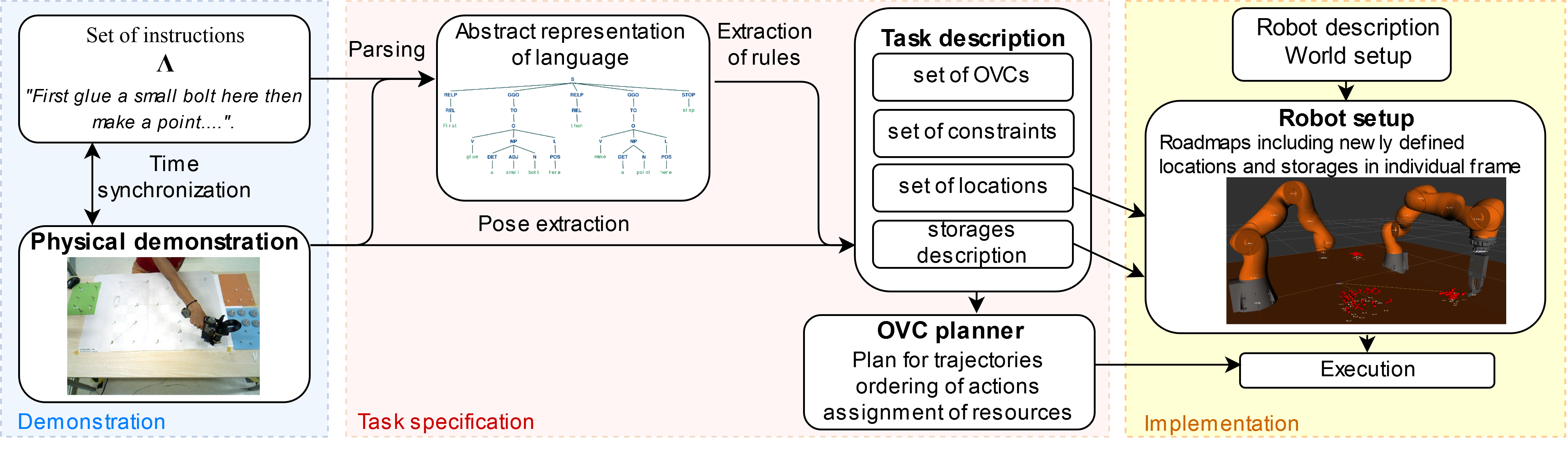}}
\caption{Overview of our architecture for imitation learning}
\label{fig:architecture}
\end{figure*}

In the following sections, we describe the challenges of capturing data for the purpose of imitation learning. We propose some solutions to the tasks based on the currently available methods and software packages. We also present our preliminary findings of the performance of the state of the art methods. The findings were gathered while implementing imitation learning architecture presented in Fig.~\ref{fig:architecture}. In Fig.~\ref{fig:setup}, we present the setup of our experiment and also manipulation tasks on which we evaluated the compared approaches.  


\section{Experimental setup}
The basic setup for the data acquisition consists of a table-desk with a calibration checkerboard, two Asus Xtion cameras with depth sensors, a high-resolution RGB camera and HTC Vive VR set. The data from all sensors are broadcasted via the Robot Operating System (ROS\footnote{http://www.ros.org/}) framework. Both Xtion sensors produce 640x480 RGB-D images at 30Hz, the RGB camera produces 5MP images at 10Hz, and HTC Vive captures the position of the controller at 60Hz. The object images, depth maps, segmentation masks and 6DoF information are extracted from raw data after calibration and time synchronization.

\section{CHALLENGES}

\subsection{Imaging Sensors}
An important part of the imitation learning setup is an imaging sensor. The choice of the sensor depends partly on the algorithms used to extract information from the scene. Some algorithms may require just a simple color camera. However, if increased positional accuracy is required, as is usually the case in industrial tasks, depth information may be necessary as well. Depth can be obtained via several types of sensors. 

\paragraph*{Stereo vision (SV)}
Stereo vision is a well-known technique for obtaining 3D information of the sensed scene~\cite{hussmann2008}. Stereo vision offers high image resolution and visual information can be obtained at the same time. The downside of stereo vision is that the field of view (FOV) is often narrow and the depth resolution is dependent on the length of the baseline, which is usually fixed. Also, a lot of computational power is required to obtain corresponding points from individual cameras, for which depth can be computed  (i.e., the correspondence problem).

\paragraph*{LiDARs}
LiDARs use laser ranging technique to obtain distance information for points in the scene. They offer high accuracy and depth resolution, long range of detection, and wide FOV. However, they are also quite expensive. Additionally, they require temporal and spatial synchronization with the visual sensor to provide RGB-D image.

\paragraph*{Time of flight cameras (ToF)}
ToF cameras use active illumination and the distance is measured from the reflected light. It can be computed either directly from the travel time of a pulse of light or indirectly from the phase shift of a modulated light~\cite{horaud2016tof}. Depth information can be calculated with much less computational power. Hence, ToF cameras can provide depth information with a higher rate than other 3D imaging technologies. They are therefore suited for application where fast moving objects might occur. However, the hardware for ToF cameras is quite expensive as high-speed electronic components are required. Moreover, the sensors usually have low image resolution.

\paragraph*{Structured light sensors (SLS)}
The biggest boom of cheap consumer 3D imaging sensors came in the form of structured light sensors. These active sensors project a structured light on the scene \cite{zanuttigh2016sls}. The depth information is then calculated from the pattern distortion caused by capturing it from a shifted viewpoint and the scene structure. The advantage of SLS cameras is their low price and ease of use. The disadvantage is lower accuracy with a relatively small working range.

There are also general considerations related to image capture. A key consideration is a computational power and transport capacity of the hardware used with the sensors. High resolution images with high update frequencies will require broad bandwidth connection. This is particularly crucial if real-time processing is required and might limit sensor placement in a real environment.

Another consideration is the ambient light. Because direct sunlight is typically many times stronger than the illumination used in active sensors, many of them (mainly ToF and SLS) are not suitable for outdoor applications. A surface of the sensed objects also plays an important role. Highly reflective or too absorptive surfaces might distort the depth measurements but may also pose a problem in object classification from RGB images. A simple solution for an undesirable surface might be masking the objects with a opaque colored tape. 

Most of the 3D imaging cameras have distance-dependent measurement accuracies. It is important to assess the working region of depth measuring, i.e., the region with sufficient depth accuracy. For most cameras, the working region starts at several dozens of centimeters. For the SV cameras, the extent of their working region depends on their geometry (baseline). The commercially available SLS cameras usually work with high enough accuracy only up to about 1 meter, with a recommended working region between 50 and 80 cm \cite{carfagni2017} even though detection range can be several meters.

One should also keep in mind that interference from multiple active sensors can occur. Some sensors can use different light frequencies or modulate the light to prevent interference. However, the performance of cheaper consumer sensors will likely be degraded in the presence of other active sensors. Acquisition speed and frame rate should also be considered. Fast moving objects captured by sensors with slow acquisition speed will result in motion blur in the color image and measurement errors in the depth image. For the gluing task the high frame rate and good image and depth resolution is more important than in the block-stacking task.

\subsection{Sensor calibration}
Calibration is an important step when preparing a visual imitation learning setup. Basically, all of the calibration procedures are based on finding corresponding points in the respective coordination systems and computing the transformation between them. The basic algorithms are usually contained in most popular software packages, such as \textit{OpenCV}\footnote{https://opencv.org/} or \textit{ROS}. Although, for higher accuracy, more sophisticated calibration should be done as these packages do not implement state of the art methods.

First, intrinsic camera parameters, such as focal length and distortion model, must be calibrated. Higher grade cameras are often supplied pre-calibrated by the manufacturer. For less expensive cameras, it is usually possible to find approximate intrinsic parameters for the specific camera model. However, it is advised to perform the calibration manually. As a result of the manufacturing process, there are always (small) differences between individual cameras even of the same make.

Extrinsic camera calibration is used to localize the camera with respect to the scene. It is necessary to correctly position the objects detected from the camera image in the scene. Extrinsic calibration is crucial when using multiple cameras and other sensors, such as LiDARs. Extrinsic parameters of individual sensors are used to fuse information captured by each sensor. When fusing information from multiple sensors, temporal synchronization is also important. Especially, when objects in the scene are expected to move at higher speeds. There are several methods to achieve temporal synchronization for imaging sensors. Authors of \cite{vsmid2017synchr} use the rolling shutter effect combined with a short burst of light while authors of \cite{albl2017synchr} use trajectory of a moving object.

Hand-eye calibration is important as well when developing and testing methods for imitation learning. Hand-eye calibration is normally used in robotics to calculate mapping from camera to robot coordinate system. In the context of imitation learning, it might be mapping from camera to the  coordinate system of a tracking device capturing the ground truth.

\subsection{Data preprocessing}
Before object detection and any other advanced processing can occur, the raw data captured by the sensors should be preprocessed. Raw image data coming from cameras should be at least rectified to remove lens distortion. Additional preprocessing may suppress effects such as motion blur. Pre-segmentation and background suppression may also be beneficial for the forthcoming processing steps.

As many methods for object classification or object pose estimation are susceptible to clutter, cropping the image as much as possible is advised. This can be done either by manually selecting the working area (e.g. crop out anything besides the working table). Alternatively, fast and robust methods for semantic segmentation or bounding box detection could be performed, such as the Mask R-CNN \cite{HeGDG17}. Afterwards, only areas believed to contain the objects of interest can be sent further down the processing pipeline. This can reduce the computational complexity and number of false positives in case of more complex scenes.

\begin{figure}[thpb]
\label{fig:interp}
\centering
\begin{tabular}{ccc}
\includegraphics[width=70 pt]{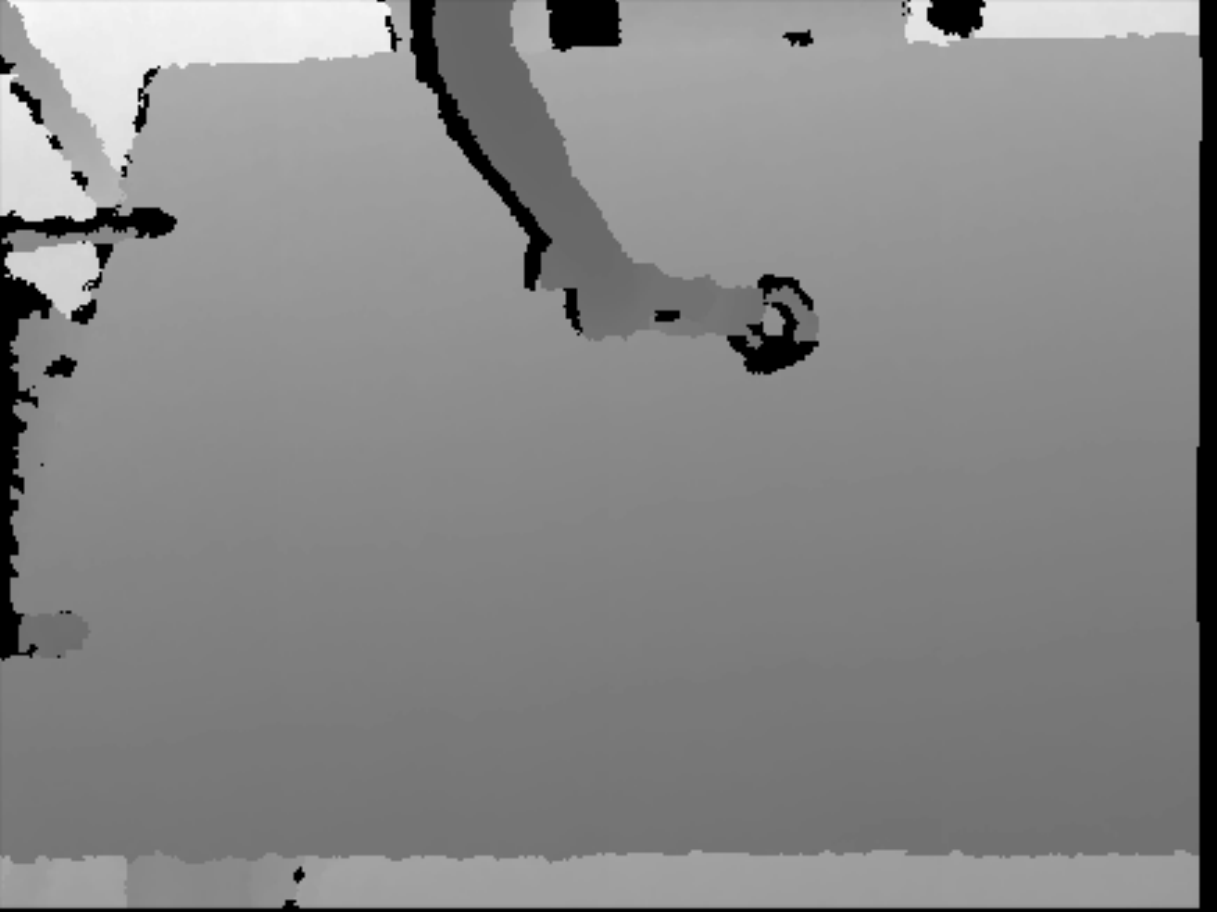} &
\includegraphics[width=70 pt]{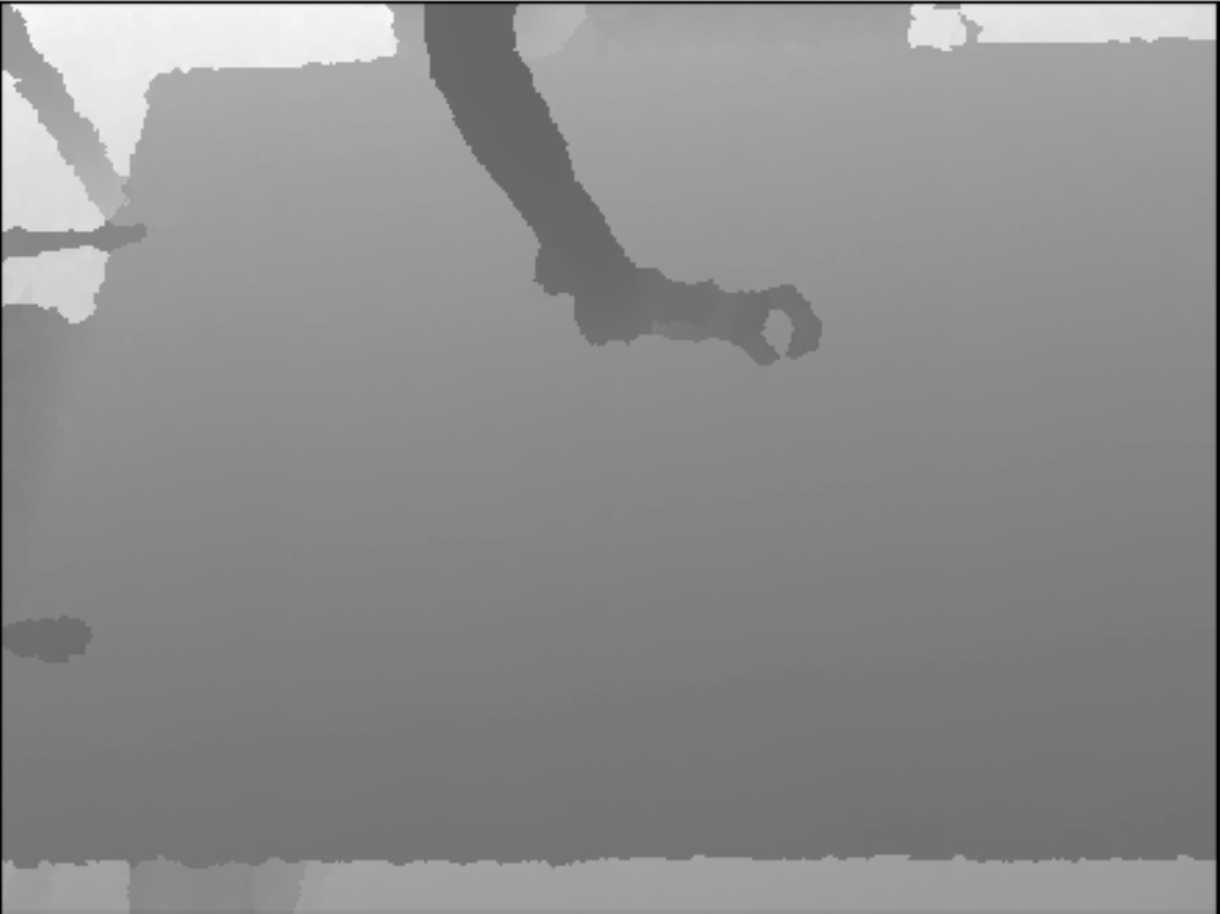} &
\includegraphics[width=70 pt]{interp_nearest.pdf}\tabularnewline
\tiny{a)} & \tiny{b)} & \tiny{c)}\tabularnewline
\end{tabular}
\caption{Depth image interpolation: a) the original image, b) depth image interpolated using nearest neighbor and c) bicubic interpolation.}
\end{figure}

Depth images can have missing depth for some pixels or contain erroneous values. These are results of either camera construction (e.g. shift of multiple cameras for SV or camera and projector for SLS) or the environment (lighting conditions, surface properties). For algorithms that can handle occlusions well enough, this might not pose any significant difficulties. However, if the used algorithms are susceptible to these artifacts, there are several methods to deal with them. One option is to use simple interpolation (see Fig.~\ref{fig:interp}). Bicubic or bilinear interpolations produces smoother gradients on the surfaces of single objects. These methods, however, also introduce unwanted effects, such as smooth transition between object and a distant background. Nearest-neighbor seems to produce better results in this case. Alternatively, more advanced smoothing can be used. For example, limiting maximum gradient of filled image patches can be used to filter out smooth transitions between objects. There are also deep network based approaches for depth image reconstruction~\cite{zhang2018}.


\subsection{Training data collection}
While gathering the training set for object detection and pose estimation, uniform sampling of the 6D pose space is important. Especially important is uniform and dense sampling in the rotational subspace; positional invariance is simpler to achieve with current machine learning algorithms. This might pose a problem if many objects are to be learned. Complex machines might be necessary to automatically position the object. Another option is to scan the object with a 3D imaging sensor and generate a 3D model which can be arbitrarily rotated. However, this method usually requires a higher-precision depth sensor to avoid errors in the model. More complex objects and surfaces with specific properties are also nearly impossible to model automatically, even with the current state of the art methods.

\begin{figure}[thpb]
\centering
{\includegraphics[width=170 pt]{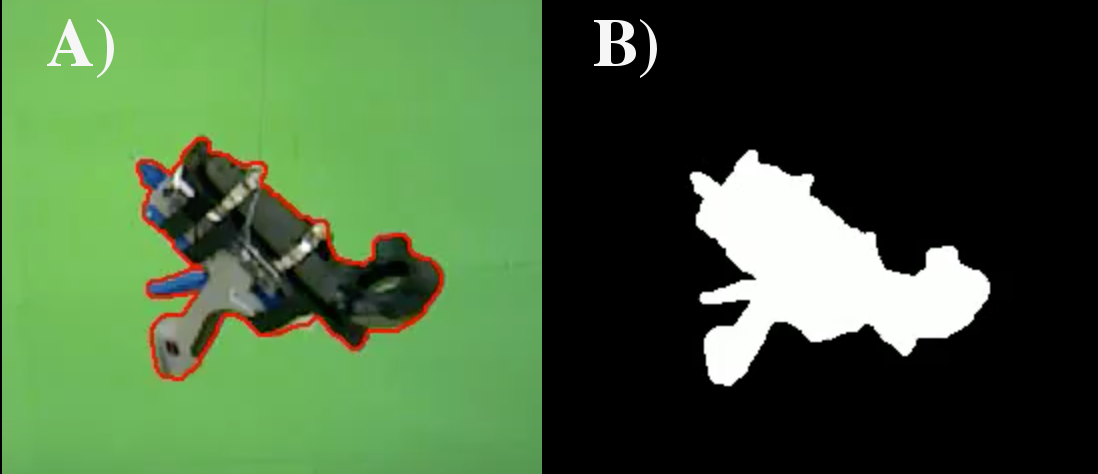}}
\caption{Sample image from the training of a used tool (gluegun). A) a gluegun in a random position on a green background (red borderline visualizes the mask boundary). B) binary mask segmented by color thresholding. The 6DOF pose information is stored in a text file for each frame.}
\label{fig:individualTool}
\end{figure}

Manual annotation of the training data is often time consuming. This can be partially solved by capturing the trained object on an easily separable background (see Fig. \ref{fig:individualTool}). To train the object on more complex backgrounds, thus gaining robustness to clutter, the images can be augmented with random backgrounds from a suitable image database. Capturing the pose ground truth can be done using traditional motion tracking techniques, such as optical motion tracking using markers. In our case, we attached the controller of a virtual reality gaming set to the tool for simple automatic 6DoF pose annotation.

\subsection{Object detection}
\begin{figure}[thpb]
\centering
{\includegraphics[width=\columnwidth]{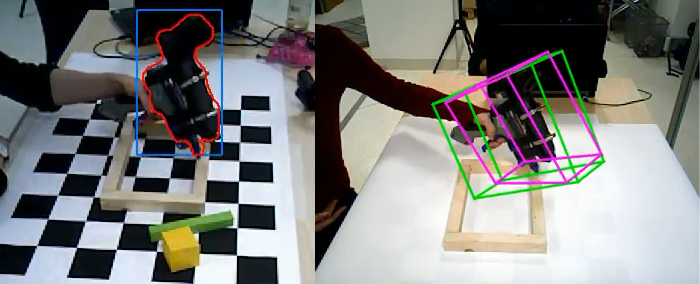}}
\caption{Detection of the tool. \textbf{Left}: MaskRCNN gluegun detection for the activity (testing) data (red shows the found segmentation of the tool, blue is the found bounding box). Note the problem when detecting the partly occluded object. \textbf{Right}: BB8 6DoF gluegun detection (green box denotes the 3D bounding box acquired from the HTC vive, purple box the 3D bounding box found by BB8 method.}
\label{fig:tool-detection}
\end{figure}

Today, object detection and classification had progressed very far. Thanks to the advent of convolutional neural networks (CNN) detections can be fast and reliable even in complex scenes. The disadvantage is that they require large amounts of training data and are computationally expensive to train. For our task, we use the Mask R-CNN algorithm~\cite{HeGDG17}. Results of the detection for the cube-stacking task can be seen in the Fig.~\ref{fig:setup}, results for the tool manipulation task can be seen in the Fig.~\ref{fig:tool-detection}. Our observation is that occlusions are problematic for the method in case of more complex objects -- in several frames, the object was not properly segmented. This might be resolved either by temporal stabilization, e.g. filtering of the position, or employing a tracking algorithm.

\subsection{6Dof pose estimation}

Motivated by applications in robotics and augmented reality, 6DoF object pose estimation has recently attracted significant attention~\cite{hodan2018}. When objects are richly textured, methods based on matching local invariant features such as SIFT \cite{munich2006sift} or SURF \cite{bay2006surf} produce reasonable results. However, in real-world and specially industrial environments objects often lack distinctive texture.
To address this issue, \cite{hodan2015detection} treat the 6DoF pose estimation from RGBD data as a (sliding window) template matching problem with different templates corresponding to different object orientations. An alternative to the template-based techniques are dense matching techniques~\cite{brachmann2014learning} 
or sparse point clouds to compute point pair features (a relative position and orientation of two points)~\cite{drost2010}. 
The point sparseness results in a much faster performance and also produces very good results in noise, clutter and partial occlusions.

In some cases, the depth information is distorted (in presence of specular materials or direct sunlight) or simply not available. Therefore it is useful to have solutions for 6DoF pose estimation from RGB images only. 
BB8 method~\cite{rad2017bb8} applies segmentation on RGB images to detect objects first in 2D and then predicts their 3D pose with Convolutional Neural Networks (CNN). This produces state-of-the-art results on the LINEMOD dataset and maintains good performance in cluttered images. An alternative is recent PoseCNN method~\cite{xiang2017posecnn}.

For the purposes of our task we utilized template matching method of Hodan et al.~\cite{hodan2015detection} and BB8 method~\cite{rad2017bb8} for RGB images. Mask R-CNN was also helpful for 2D shape detection for the BB8 method; the results for 6D pose estimation by BB8 (without refinement) were rather poor, however (especially in presence of occlusion). Preliminary results can be seen in Fig.~\ref{fig:tool-detection}. 


In our experiments, we encountered high sensitivity to the quality of the depth data and its synchronization with the image data. This is the main problem in the case of the quick movements of the objects, i.e., for the tool manipulation tasks, since even a small distortion in the 6DoF pose estimation of the whole object will cause a significant position error of the tip of the tool. Especially for Hodan method we observed a significant improvement in detections when we used segmentation from Mask R-CNN to cut region of interest from the image. 
Mask R-CNN was also helpful for 2D shape detection for the BB8 method; the results for 6D pose estimation by BB8 (without refinement) were rather poor, however (especially in presence of occlusion, as can be seen in~Fig.~\ref{fig:tool-detection}). On top of these, as known, BB8 method has problems with symmetrical objects. In the stacking task, the detections were more reliable as the manipulated objects are simple geometric shapes.

Furthermore, it is worth noting that datasets on which are these 6DoF methods typically evaluated are very different from data which are important for imitation learning. Typical dataset includes only static objects, fixed distance of camera from the object, similar distance of camera from objects in training and testing dataset, ideal environment with ideal lightning conditions and well synchronized depth and RGB images.



\subsection{Tracking human pose}

For imitation of a human demonstrated activity, tracking human body is an obvious task -- e.g., human joints in space. Estimation of human joints based on RGBD sensor is usually referenced as human pose estimation, skeleton tracking or skeletal pose. Current methods estimate human pose from depth images or monocular images or join depth and image data~\cite{zimmermann20183d}. 

The state-of-the-art challenge is to estimate human pose while holding some a priori unknown objects or occlusions of body parts. In \cite{yao2010modeling}, the authors proposed a method for taking advantage of a context of human body parts and observed scene. Unfortunately, it still remains unsolved.

We observed ROS compatible open-source tools for estimation of  human pose which is OpenNI and NiTE. Given an RGB-D image, we can process it in $openni tracker$ package\footnote{$http://wiki.ros.org/openni\_tracker$} to obtain 15 transformations to the specified world for each human estimated joint. It is namely: \textit{head, neck, torso, left shoulder, left elbow, left hand, right shoulder, right elbow, right hand, left hip, left knee, left foot, right hip, right knee, right foot}. These transformations are combined together as a transformation network which represents human pose in terms of joints for recording his/her actions.


\begin{figure}[thpb]
\centering
{\includegraphics[width=120 pt]{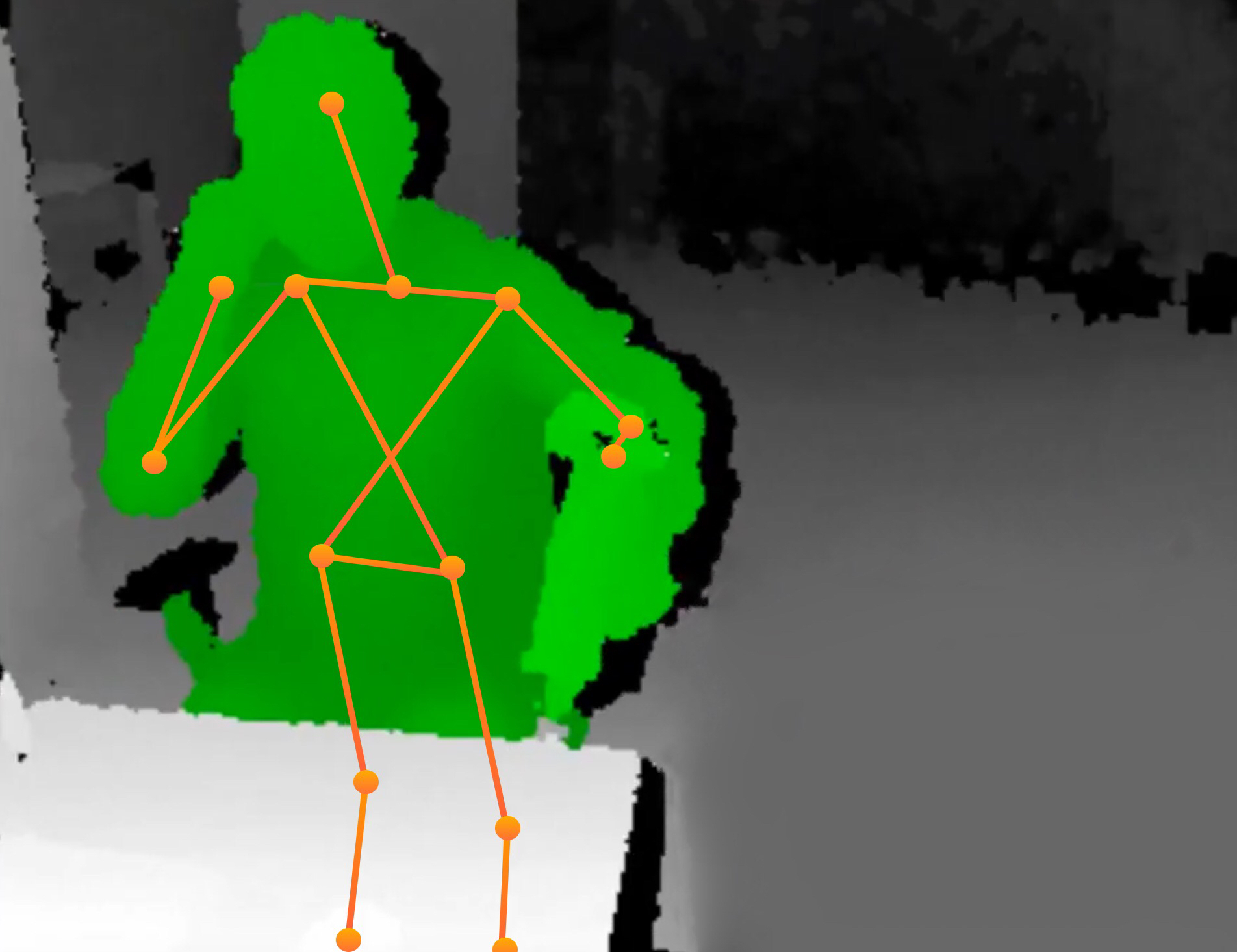}}
\caption{The failure of the human pose estimation algorithm to correctly estimate the pose of a sitting person that scratches their head while holding an object in the other hand. Note that the occlusion of legs behind the table causes a wrong estimate of the lower body parts. In the left arm the human has a tool which causes a wrong estimation of the wrist joint.}
\label{fig:rgbdFail}
\end{figure}

While holding an object in hands during manipulation tasks, we face huge variability of estimation, which leads to instability and unreliability of the estimation. Detection fails (see Fig.~\ref{fig:rgbdFail})
mainly in occlusions and atypical positions (such as couching, sitting behind desk or putting hand into pocket). Many of the standard methods use pre-trained models trained on typical position of standing human in good light conditions. In any other position segmentation starts to fail and provide imprecise estimates of joint position and orientation. Another limitation is correct tracking of multiple humans, especially if overlap occurs or a person leaves the scene temporarily. However, there are recent attempts to detect human pose in wild. For example, Papandreou et al.~\cite{papandreou2017} proposed a two stage method for multi-person detection and 2-D pose estimation in wild from RGB images.

 
OpenPose is an alternative algorithm~\cite{cao2017realtime} which uses only RGB information and estimates a maximum of 25 joints and 70 key point face points. Since this method uses only RGB information, only 2D information can be inferred and additional mapping to depth image is needed. 



\subsection{Language}

Language commands can serve as a useful supplementary source of information for cases where the robot cannot extract enough data from vision -- e.g., detecting time and pose of contact, exact position such as a corner, etc.. However, mapping between vision and language brings up a number of constraints which need to be considered. 

Firstly, a method for language grounding needs to be adopted, so that the robot is able to transform the command into its contextual perception. Semantic parsing, i.e. mapping between a natural language (NL) sentence and its logical representation, can be obtained using a probabilistic Combinatorial Categorical Grammar (CCG) \cite{kwiatkowski2010inducing}. However, this approach relies on manually defined rules for such mapping - therefore, for our task it is more suitable to adopt a method where the relations are learned from data or human dialogues and can be applied for different environments. Such methods have been proposed e.g. in \cite{matuszek2013learning}\cite{thomason2015learning}\cite{alomari2017natural}.

Secondly, it is important to obtain accurate temporal synchronization between performed actions and spoken commands. This must be held in mind by the person who is demonstrating the task and thus makes the framework less user-friendly. This issue can be minimized in final stages of the model development, e.g. by using the dependency relation matrix proposed in \cite{cheng2017modeling}, enabling the model to extract correct command sequences from wrongly ordered inputs.

To implement language in our framework, we have been progressing in a bottom-up direction. The first goal is selecting a suitable automatic speech recognition (ASR) system. We have compared between the open-source version of Google Speech API and CMUSphinx Open Source Speech Recognition system. In CMUSphinx, we compared between a naive, untrained version and an adapted acoustic model with custom ARPA language model, specific to our gluegun task. Testing of all three systems (Google Speech API, default and adapted CMUSphinx) showed that for a predefined task with a limited vocabulary, it is most plausible to train a custom model with CMUSphinx (the measured WER for the trained model was below 1\%). However, if we cannot specify the vocabulary beforehand, Google Speech API has a better general performance (16\% WER compared to 76\% WER for untrained CMUSphinx). Therefore, our selected ASR system for the imitation learning task was Google Speech API. 




\section{CONCLUSION}

Enabling learning only by visual demonstration without any on-teacher sensors faces many challenges and opportunities. In this paper, we highlighted them and proposed the possible solutions. There are already new emerging algorithms which try to deal with many of these challenges (such as human pose detection in occlusions and atypical positions, etc.). However, we see the way to the cognitive robot, capable of understanding and imitating human activity without any external sensors and markers, still as a challenging task. Many of the current machine learning algorithms will have to be adopted and improved and novel approaches will have to be developed. In our future work, we want to use the presented findings and incorporate them to our architecture. We would like to quantitatively evaluate RGB and RGB-D methods (namely \cite{hodan2018} and \cite{xiang2017posecnn} methods) in our demonstration tasks.





\addtolength{\textheight}{-12cm}   




\section*{ACKNOWLEDGMENT}
This work was supported by the European Regional Development Fund under projects Robotics for Industry 4.0 (reg. no. $CZ.02.1.01/0.0/0.0/15\_003/0000470$), TA\v{C}R Zeta project Imitation learning supported by language for industrial robots  (no. TJ01000470) granted by Technological agency of Czech republic, and IMPACT (reg. no. $CZ.02.1.01/0.0/0.0/15\_003/0000468$).

\bibliographystyle{IEEEtran}
\bibliography{iros-workshop}

\end{document}